\documentclass[conference]{IEEEtran}

\ifCLASSINFOpdf
\else
\fi

\IEEEoverridecommandlockouts
\usepackage[x11names]{xcolor}
\usepackage{lscape}
\usepackage{graphicx}
\usepackage{booktabs}
\usepackage{colortbl}
\usepackage{pifont}
\usepackage{amssymb}
\usepackage[inkscapelatex=false]{svg}
\usepackage{tablefootnote}
\usepackage{tikz}
\usepackage{url}
\usepackage{amsmath}
\usepackage{enumitem}
\usepackage{multirow}
\usepackage{longtable}
\usepackage{booktabs}
\usepackage{silence}
\usepackage[most]{tcolorbox}
\usepackage{tabularray}
\usepackage{colortbl}
\usepackage{hhline}
\usepackage{pdfpages}

\setlist[itemize]{topsep=1.65mm}

\usepackage[none]{hyphenat}

\usepackage{array} 
\newcolumntype{P}[1]{>{\centering\arraybackslash}p{#1}}
\newcolumntype{M}[1]{>{\centering\arraybackslash}m{#1}}

\usepackage{url}

\usepackage{colortbl}

\tcbset{highlight math style={colback=Ivory1!0, colframe=Snow1!0, boxrule=0.5pt, coltext=black}}

\definecolor{customblue}{HTML}{330099} 
        
\newcommand{\xmark}{\ding{53}}  

\vbadness=\maxdimen

\graphicspath{ {./images/} }

\title{A Framework for Measuring the Quality of Infrastructure-as-Code Scripts}
\author{
    \IEEEauthorblockN{Pandu Ranga Reddy Konala\IEEEauthorrefmark{2}, Vimal Kumar\IEEEauthorrefmark{2}, David Bainbridge\IEEEauthorrefmark{2}, Junaid Haseeb\IEEEauthorrefmark{2}}
    \IEEEauthorblockA{\IEEEauthorrefmark{2}School of Computing and Mathematical Sciences, University of Waikato\\Hamilton, New Zealand 3240
    \\\{pkonala, vkumar, davidb, jhaseeb\}@waikato.ac.nz}
}

\begin{document}

\maketitle

\begin{abstract} 
Infrastructure as Code (IaC) has become integral to modern software development, enabling automated and consistent configuration of computing environments. The rapid proliferation of IaC scripts has highlighted the need for better code quality assessment methods. This paper proposes a new IaC code quality framework specifically showcased for Ansible repositories as a foundation. By analyzing a comprehensive dataset of repositories from Ansible Galaxy, we applied our framework to evaluate code quality across multiple attributes. The analysis of our code quality metrics applied to Ansible Galaxy repositories reveal trends over time indicating improvements in areas such as metadata and error handling, while highlighting declines in others such as sophistication and automation. The framework offers practitioners a systematic tool for assessing and enhancing IaC scripts, fostering standardization and facilitating continuous improvement. It also provides a standardized foundation for further work into IaC code quality.
\end{abstract}




\section{INTRODUCTION}

Infrastructure as Code (IaC) has revolutionized IT infrastructure management by treating configurations as version-controlled, executable code, enabling automation, scalability, and consistency. However, as IaC becomes integral to software development, specific quality attributes are needed, since traditional standards like  ISO/IEC 25010:2011 (Systems and software Quality Requirements and Evaluation — SQuaRE) \cite{2013ISOI} are designed for application software and do not fully address the unique characteristics of IaC.

In this paper, we introduce an IaC quality framework consisting of nine categories, as presented herein, aligned with the aforementioned ISO/IEC 25010:2011 standard. This framework builds on and differentiates from prior work by Rahman and Williams \cite{Rahman_2019}, Dalla Palma et al.\ \cite{Palma2020TowardsAC}, and Begoug et al.\ \cite{TerraMetrics} by tailoring the categories to the specific context of IaC. These categories—Metadata, Code Structure, Code Sophistication, Code Maintainability, Code Security, Function and Purpose, Error Handling, Automation, and Code Integration—provide a comprehensive framework that encompasses all identified IaC code attributes while remaining grounded in established principles.

We validated the framework through empirical analysis of repositories from Ansible Galaxy \cite{AnsibleGalaxy}, which highlighted notable trends in IaC script quality.
This analysis affirms the relevance of the proposed categories and demonstrates its practical applicability in evaluating real-world IaC scripts. 

 We present this investigation into the relationship between the attributes and IaC quality assessment practices in the form of five key research questions (RQs).
 
   
\begin{itemize}

     \item[\scalebox{0.75}{\ding{108}}] \textbf{RQ1:} \textit{What mechanisms are currently employed by IaC tools to evaluate the quality of their scripts, and how consistent are these methods across the different tools?}
    \\
    \item[\scalebox{0.75}{\ding{108}}] \textbf{RQ2:} \textit{How do existing IaC code quality attributes map to the characteristics defined in the ISO/IEC 25010:2011 standard, and what are the limitations of this mapping?}
    \\
    \item[\scalebox{0.75}{\ding{108}}] \textbf{RQ3:} \textit{Which specific IaC code quality categories can be developed to accommodate all identified IaC code attributes while drawing inspiration from ISO/IEC 25010:2011?}
    \\
    \item[\scalebox{0.75}{\ding{108}}] \textbf{RQ4:} \textit{How does the proposed IaC code quality framework perform when applied to real-world IaC repositories, and what insights can be gained?}
    \\
    \item[\scalebox{0.75}{\ding{108}}] \textbf{RQ5:} \textit{In what ways can the proposed IaC code quality framework enhance the assessment of IaC scripts in practice?}
\end{itemize}
\section{BACKGROUND}

This section establishes a foundation for analyzing IaC quality, focusing on tools for managing IT infrastructures. It compares IaC tools such as Ansible, Terraform, Chef, and Puppet, examining their features and assessing their suitability in various contexts. By evaluating their strengths and limitations, the section provides insights into tool usage and discusses methodologies for code quality evaluation and assurance. It also compares frameworks and scoring mechanisms to explore their influence on IaC quality management.

IaC is categorized into three areas \cite{Wang_2022} that automate IT processes: \textit{Infrastructure Provisioning} automates allocation and management of hardware resources, with tools like Terraform and AWS CloudFormation \cite{CloudFormation} enabling resource deployment. \textit{Configuration Management} automates software infrastructure configuration; tools such as Ansible, Chef, and Puppet codify system configurations, reducing errors and enhancing reproducibility. \textit{Image Building} standardizes machine and container images; tools like Packer \cite{PackerBook} create machine images, while Docker \cite{DockerBook} enables containerization for application portability. In summary, IaC tools automate provisioning, configuration, and image building, improving efficiency and reliability in IT operations.

\subsection{COMPARING EXISTING IaC TOOLS AND PRACTICES}
\par
\scalebox{0.75}{\ding{108}} \textbf{RQ1:} \textit{What mechanisms are currently employed by IaC tools to evaluate the quality of their scripts, and how consistent are these methods across the different tools?}
\\

IaC tools employ mostly rudimentary mechanisms to evaluate script quality, but these mechanisms also differ across platforms and categories, indicating a lack of standard assessment methodologies as shown in Table~\ref{table1}. For instance, repositories such as Ansible Galaxy, Puppet Forge, and Chef Supermarket implement quality control measures, but their implementation and transparency vary. Configuration management tools incorporate custom scoring systems to ensure quality control in maintaining IaC repositories. Notably, infrastructure provisioning and image building tools typically lack these mechanisms or have assessments that are less common.

\begin{table*}[ht!]
\centering
\caption{IaC Software Tools Repository \& Quality Scoring Metrics}
\label{table1}
\begin{tblr}{
  rowsep = 0pt,
  row{1} = {SlateGray4!35},
  cell{2}{1} = {r=4}{SlateGray4!15},
  cell{6}{1} = {r=4}{SlateGray4!15},
  cell{10}{1} = {r=2}{SlateGray4!15},
  cell{2-13}{2} = {Snow1!20},
  cell{2-13}{3} = {Snow1!20},
  cell{2-13}{4} = {Snow1!20},
  vlines,
  hline{1} = {-}{0.4mm},
  hline{last} = {-}{0.4mm},
  vline{1} = {0.4mm},
  vline{last} = {0.4mm},
  vline{5} = {0.4mm},
  vline{2} = {0.3mm},
  vline{3} = {0.3mm},
  vline{4} = {0.3mm},
  hline{1-2,6,10,12} = {-}{0.3mm},   
  hline{3-5,7-9,11} = {2-4}{0.1mm},  
}
\SetCell{c}\textbf{IaC Categories}              & \SetCell{c}{\textbf{IaC Software Tools}}                                                          & \SetCell{c}\textbf{Repository}                                                         & \textbf{Quality Scoring Metrics}       \\
\SetCell{c}\textbf{Infrastructure Provisioning} & \SetCell{c}Terraform \cite{TerraformBook}                                      & \SetCell{c}Terraform Registry \cite{TerraformRegistry}                & \SetCell{c}None                                   \\
                                     & \SetCell{c}CloudFormation (AWS)                                                                 & \SetCell{c}None                                                                        & \SetCell{c}None                                   \\
                                     & \SetCell{c}Azure Resource Manager (ARM) Templates \cite{AzureARMTemplates}     & \SetCell{c}Azure Quickstart Templates \cite{AzureQuickstartTemplates} & \SetCell{c}Pass/Fail                              \\
                                     & \SetCell{c}Google Cloud Deployment Manager \cite{GoogleCloudDeploymentManager} & \SetCell{c}None                                                                        & \SetCell{c}None                                   \\
\SetCell{c}\textbf{Configuration Management}    & \SetCell{c}Ansible \cite{AnsibleBook}                                          & \SetCell{c}Ansible Galaxy                                                              & \SetCell{c}Scale of 0-5 points                    \\
                                     & \SetCell{c}Puppet \cite{PuppetBook}                                            & \SetCell{c}Puppet Forge \cite{PuppetForge}                            & \SetCell{c}Scale of 0-100\%                       \\
                                     & \SetCell{c}Chef \cite{ChefBook}                                                & \SetCell{c}Chef Supermarket                                                            & \SetCell{c}Scale of 0-100\%                       \\
                                     & \SetCell{c}SaltStack (Salt) \cite{SaltStack}                                   & \SetCell{c}SaltStack Formula \cite{SaltStackFormula}                  & \SetCell{c}None                                   \\
\SetCell{c}\textbf{Image Building}              & \SetCell{c}Packer                                                                               & \SetCell{c}None                                                                        & \SetCell{c}None                                   \\
                                     & \SetCell{c}Docker                                                                               & \SetCell{c}Docker                                                                      & \SetCell{c}Verified Publisher Program 
\end{tblr}
\end{table*}

The analysis of quality scoring systems across various IaC platforms, detailed in Table~\ref{table:quality_scoring_comparison}, reveals heterogeneity in assessment practices and a lack of collective standardization. This variation poses challenges for comparability and consistency, potentially impacting user trust and assessment reliability. This highlights the need for standardized, transparent, and comprehensive quality metrics for IaC tools. Implementing a unified quality assessment approach would facilitate informed decision-making and enhance the reliability and efficiency of the IaC ecosystem.

\begin{table*}[ht!]
\centering
\caption{Comparison of Quality Scoring Systems across Ansible Galaxy, Puppet Forge, Chef Supermarket, Azure, and Docker}
\label{table:quality_scoring_comparison}
\begin{tabular}{|M{2cm}|M{2.5cm}|M{2.5cm}|M{2.5cm}|M{2.5cm}|M{2.5cm}|}
\hline
\rowcolor{SlateGray4!35}\textbf{Feature} & \textbf{Ansible Galaxy} & \textbf{Puppet Forge} & \textbf{Chef Supermarket} & \textbf{Azure} & \textbf{Docker} \\
\hline
\cellcolor{SlateGray4!15}\textbf{Scoring Method} & \cellcolor{Snow1!20}Weighted combination of community and quality scores, based on user surveys. & \cellcolor{Snow1!20}Automated evaluation for code formatting and malware checks. & \cellcolor{Snow1!20}Community-defined quality metrics, including publication status and platform support. & \cellcolor{Snow1!20}Compliance checks against Azure best practices. & \cellcolor{Snow1!20}Docker Security Scanning and Docker Verified Publisher Program. \\
\hline
\cellcolor{SlateGray4!15}\textbf{Scoring Scale} & \cellcolor{Snow1!20}Point based scoring ranging from 0 to 5.0 for both Community and Quality Scores. & \cellcolor{Snow1!20}Percentage based scoring. & \cellcolor{Snow1!20}Percentage based scoring. & \cellcolor{Snow1!20}Binary Pass/Fail assessment. & \cellcolor{Snow1!20}None \\
\hline
\cellcolor{SlateGray4!15}\textbf{Evaluation Tools} & \cellcolor{Snow1!20}{\em yamllint} \cite{Yamllint}, {\em ansible-lint} \cite{AnsibleLint}, import-time checks. & \cellcolor{Snow1!20}VirusTotal for binary security checks. & \cellcolor{Snow1!20}Foodcritic \cite{8590193} for cookbook evaluation. & \cellcolor{Snow1!20}Azure Policy, ARM Template Test Toolkit. & \cellcolor{Snow1!20}Docker Scout for vulnerability analysis. \\
\hline
\cellcolor{SlateGray4!15}\textbf{Score Components} & \cellcolor{Snow1!20}Community input, Syntax Score, Metadata Score. & \cellcolor{Snow1!20}Code formatting, security checks. & \cellcolor{Snow1!20}Publication status, collaboration, licensing, platform support. & \cellcolor{Snow1!20}Compliance with Azure standards, security checks, resource optimization. & \cellcolor{Snow1!20}Security vulnerabilities, image integrity, publisher verification. \\
\hline
\cellcolor{SlateGray4!15}\textbf{Critique Points} & \cellcolor{Snow1!20}User survey dependence, temporal limitation, tool efficacy. & \cellcolor{Snow1!20}Lack of detailed documentation, focus on code and malware only. & \cellcolor{Snow1!20}Emphasis on objective metrics, late-stage evaluation, community feedback reliance. & \cellcolor{Snow1!20}Limited visibility into detailed scoring criteria. & \cellcolor{Snow1!20}Limited visibility into detailed scoring criteria. \\
\hline
\end{tabular}

\centering
\end{table*}

\subsection{RELATED WORK}

IaC has transformed the management and provisioning of IT infrastructure; however, maintaining the quality and standardization of IaC scripts continues to be a critical challenge. A survey by Konala et al.\ \cite{pandusok} highlights that several tools exist that are designed to address specific quality issues such as security smells or code smells through the use of various techniques like regular expressions, linting, deep learning, etc. Despite the advancements made with these tools, our analysis reveals a lack of  quality assessment approaches to evaluating overall IaC code quality. This gap underscores the need for a higher-level solutions that evaluate IaC scripts.

Most existing work addresses this research gap by 
proposing quality attributes and metrics tailored to different IaC tools. Rahman et al.\ \cite{Rahman_2019} initially conducted an empirical study on Puppet scripts, identifying 12~source code properties correlated with defects, such as lines of code and hard-coded strings. Their defect prediction models achieved precision between 0.70--0.78 and recall between 0.54--0.67, offering practitioners actionable insights to enhance script quality. However, the study’s focus on Puppet scripts limits its applicability to other IaC tools like Ansible, Chef, or Terraform. Additionally, the lack of alignment with ISO standards suggests the need for work to make these attributes more universally applicable.

Similarly, Dalla Palma et al.\ \cite{Palma2020TowardsAC} developed a catalog of 46 quality attributes specific to Ansible scripts, addressing aspects such as code complexity, maintainability, error handling, and best practices. While beneficial for Ansible users, these attributes need adaptation and validation to apply across other IaC environments like Chef or Terraform. Additionally, their impact on quality outcomes has yet to be empirically validated. Begoug et al.\ introduced TerraMetrics \cite{TerraMetrics}, a tool for measuring Terraform script quality using 40 attributes based on static code analysis which draws parallels to Dalla Palma et al.\ \cite{Palma2020TowardsAC} study. Despite its focus on code quality, its limited validation on only three repositories highlights the need for broader assessment. Overall, the previous work in this area points to a need for standardized quality assessment that can be applied across diverse IaC tools.

\begin{tcolorbox}[
    colframe=SlateGray4, 
    colback=Snow1!20,  
    boxrule=1pt, 
    rounded corners
]
\ding{228} \textbf{RQ1 Takeaway Message:} \textit{IaC tools use wide range of methods to evaluate script quality, creating challenges in ensuring consistency and comparability across platforms.}
\end{tcolorbox}

\subsection{ABOUT ISO/IEC 25010:2011 SOFTWARE QUALITY STANDARDS}

ISO/IEC 25010:2011, part of the Systems and Software Quality Requirements and Evaluation (SQuaRE) model, is an international standard for assessing software and system quality. It builds on the ISO/IEC 9126 \cite{ISO9126} standard by introducing ten quality characteristics that help organizations evaluate key aspects such as functionality and reliability. The framework consists of two models: the Quality in Use Model, which includes five characteristics related to user interaction applicable to human-computer interfaces, and the Product Quality Model, which encompasses eight characteristics addressing the static and dynamic aspects of software and computer systems. In the context of IaC and this research, the Product Quality Model is particularly relevant. IaC involves managing and provisioning infrastructure through machine-readable definition files, which aligns with the static and dynamic properties covered by the model’s eight characteristics. Applying the Product Quality Model to IaC ensures that infrastructure provisioning meets operational and quality standards.


In terms of relating the work presented here to ISO/IEC 25010:2023 \cite{ISO25010:2023}---the relevant part of the successor to ISO/IEC 25010:2011, but at the time of writing yet to confirmed as a standard---there are two key areas of change.  The first is that ``usability'' has been renamed to ``interaction capability'' and ``portability'' to ``flexibility,'' which are cosmetic changes. The second is the inclusion of an additional characteristic: ``safety.''  This covers aspects such as failsafe mechanisms and operational constraints.  
As current IaC tools provide limited support for this, which itself is through 3rd-party plugins, this was not included in the analysis reported here.  
We note however, that it would be a straightforward matter to reapply the methodology 
to include this additional characteristic, in response to safety feature support becoming mainstream in IaC tools.

\subsection{RESEARCH GAPS IN STATE-OF-THE-ART STUDIES}

Despite advancements in assessing IaC quality, significant gaps persist due to the lack of standardized calculations and theoretical foundations. Much of the existing research is tool-specific, resulting in fragmented approaches that limit the generalizability of findings across different IaC platforms. This fragmentation hinders the development of a robust framework for evaluating IaC code quality, which is essential for ensuring consistent assessments.

Rahman et al.'s \cite{Rahman_2019} study on Puppet scripts, for instance, highlights code properties linked to defects but is constrained by its narrow focus on Puppet and its non-alignment with international software quality standards like ISO/IEC 25010:2011. Similarly, Dalla Palma et al.'s \cite{Palma2020TowardsAC} work on Ansible scripts catalogues 46 quality attributes but lacks empirical validation and standardization like Rahman et al.'s work. Tools like TerraMetrics \cite{TerraMetrics}, designed for Terraform, also suffer from limited validation, small sample sizes, and non-adherence to recognized standards, raising concerns about their broader applicability.

Overall these approaches primarily focus on identifying low-level issues such as security smells
or code smells without providing an evaluation of
overall code quality. Several IaC tool repositories provide their own quality scoring mechanisms. The lack of thorough documentation about these mechanisms however, creates a transparency issue in addition to the issue of consistency as each platform employs its own mechanism (see Table \ref{table:quality_scoring_comparison}). This underscores the need for a
standardized and empirically-backed framework, to improve IaC quality assessment and support the reliability of modern IT infrastructures.


\section{METHODOLOGY FOR MAPPING IaC CODE ATTRIBUTES}

In IaC, attributes serve as fundamental components of configuration scripts. Defined as key-value pairs, these attributes convey information about configurations, target systems, permissions, and conditions necessary for infrastructure setup and management. Attributes direct the execution of IaC scripts, detailing the specific tasks to be performed, on which systems, and with what permissions. 
Analyzing such attributes for their alignment with quality characteristics, provides a basis for assessing their suitability in meeting quality requirements.

To illustrate, consider a basic Ansible playbook snippet \cite{AnsibleBook} that installs Nginx:

\begin{tcolorbox}[
    colframe=SlateGray4,        
    colback=Snow1!20,     
    boxrule=1pt,           
    rounded corners,        
    title={Example Ansible Script for Installing Nginx},
    halign title=flush center,
    parbox=false,
    enhanced,
]
\begin{lstlisting}[]
- name: Install Nginx
  hosts: web_servers
  become: yes
  tasks:                       
    - name: Install Nginx
      apt:
        name: nginx
        state: present
\end{lstlisting}
\end{tcolorbox}

In the above example, each attribute is described based on definitions from the Ansible documentation \cite{Ansible}, contextualized for this specific case:

\begin{itemize}
    \item The \textcolor{customblue}{\texttt{\textbf{name}}} attribute specifies the purpose of the play, here defined as ``Install Nginx.''
    \item \textcolor{customblue}{\texttt{\textbf{hosts}}} designates the target machines for task execution, in this case, a group labeled \texttt{web\_servers}.
    \item \textcolor{customblue}{\texttt{\textbf{become}}} enables privilege escalation, allowing the task to execute with elevated permissions.
    \item \textcolor{customblue}{\texttt{\textbf{tasks}}} lists actions to perform on the target servers. Each entry in the \textcolor{customblue}{\texttt{\textbf{tasks}}} list represents a specific task that Ansible will carry out.
    \item \textcolor{customblue}{\texttt{\textbf{apt}}} specifies the module used in this task, in this case, managing packages on Debian-based systems.
    \item \textcolor{customblue}{\texttt{\textbf{state}}} is a key within the \textcolor{customblue}{\texttt{\textbf{apt}}} module that defines the desired state of the package. Setting \textcolor{customblue}{\texttt{\textbf{state}}}\texttt{: present} ensures that Nginx will be installed if it is not already on the system.
\end{itemize}

These attributes define the operational parameters of the playbook, establishing its scope, permissions, and task requirements. The usage of attributes in a script such as the one above can be mapped to quality characteristics such as in ISO/IEC 25010:2011.
Building on this foundation, the following section describes the methodology developed to address code quality issues in IaC scripts, structured into three key steps.
\begin{itemize}
\item \textit{Step 1: Mapping IaC Code Attributes to ISO/IEC 25010:2011.} This initial mapping examines the extent to which the ISO/IEC 25010:2011 standard can accommodate specific IaC code attributes.
\\
\item \textit{Step 2: Development and Mapping of IaC Code Attributes to the Proposed IaC Code Quality Framework.} This step assesses how the proposed framework meets the unique requirements of IaC, especially where ISO/IEC 25010:2011 may have limited applicability.
\\
\item \textit{Step 3: Comparison of Mapping Between ISO/IEC 25010:2011 and the Proposed IaC Code Quality Framework.} This final step presents a comparative analysis, demonstrating the proposed framework’s capacity to capture IaC-specific characteristics in comparison with the ISO/IEC 25010:2011 standard.
\end{itemize}

\subsection*{STEP 1: MAPPING IaC CODE ATTRIBUTES TO ISO/IEC 25010:2011}
\par
\scalebox{0.75}{\ding{108}} \textbf{RQ2:} \textit{How do existing IaC code quality attributes map to the characteristics defined in the ISO/IEC 25010:2011 standard, and what are the limitations of this mapping?}
\\

Building on Rahman et al.'s \cite{Rahman_2019} identification of 12 source code properties/attributes indicating defects in Puppet scripts and Dalla Palma et al.'s \cite{Palma2020TowardsAC} expansion to 46 attributes in Ansible scripts, we assesses whether the ISO/IEC 25010:2011 framework is able to capture IaC's quality features. The quality attributes from previous work were categorized relative to ISO/IEC 25010:2011 based on their definitions into \textit{direct mapping}, \textit{indirect mapping}, and \textit{no mapping}. Direct mapping includes IaC properties that correspond to ISO sub-characteristics and attributes, indirect mapping includes attributes that align operationally with ISO sub-characteristics without direct correspondence, and no mapping includes attributes unique to IaC that do not fit within the ISO framework.

During the mapping source properties identified by Rahman et al., attributes such as ``Hard-Coded String'' and ``Lines of Code'' were \text{indirectly mapped} to ISO/IEC 25010:2011 due to their impact on security and portability, respectively, while the ``URL'' metric had \text{no mapping} as it does not align with the standard's quality characteristics. Table \ref{mapping_iso_rahman}, shows the complete mapping. Similarly, Dalla Palma et al.’s analysis of 46 Ansible script attributes features like ``Variables,'' ``Regex,'' ``Conditions,'' ``Decisions,'' and ``User Interactions'' as \text{indirectly mapped} to ISO characteristics such as functional suitability, performance efficiency, and usability, and attributes like ``Suspicious Comments'' and ``Deprecated Keywords and Modules'' were linked to security concerns. Attributes such as ``Plays,'' ``Roles,'' and ``Filters'' had \text{no direct} mapping due to their unique roles within IaC. Table \ref{mapping iso dalla palma} shows the complete mapping. 

In both tables it can be seen that there are source properties that cannot be mapped to any ISO characteristic. As the number of properties increase in a given quality measurement system, the number of properties that can't be mapped to any ISO characteristic will also increase.  This demonstrates the limitations of the ISO standard in fully capturing IaC-specific quality attributes and highlights the need for a framework tailored to evaluate IaC code quality effectively. 
It should be noted that Begoug et al.’s \cite{TerraMetrics} work was excluded from this mapping given their alignment with Dalla Palma et al.\cite{Palma2020TowardsAC} for simplicity.

\begin{tcolorbox}[
    colframe=SlateGray4, 
    colback=Snow1!20,  
    boxrule=1pt, 
    rounded corners
]
\ding{228} \textbf{RQ2 Takeaway Message:} \textit{Although certain IaC code quality attributes correspond directly or indirectly to the characteristics of ISO/IEC 25010:2011, many fall outside its scope, revealing limitations in the standard's ability to capture IaC-specific code quality features.}
\end{tcolorbox}

\begin{table}[!ht]
    \centering
    \caption{Mapping of IaC Code Attributes of Rahman et al.'s \cite{Rahman_2019} study to ISO/IEC 25010:2011}
    \label{mapping_iso_rahman}
    \begin{tabular}{|M{2.75cm}|M{2.5cm}|M{2.25cm}|}
    \hline
    \rowcolor{SlateGray4!35} \textbf{\textbf{\begin{tabular}[c]{@{}l@{}}ISO 25010:2011\\Characteristics\end{tabular}}} & \textbf{Direct Mapping} & \textbf{Indirect Mapping} \\ \hline
    \cellcolor{SlateGray4!15}\textbf{Functional Suitability} & \cellcolor{Snow1!20}Include, Ensure, Require, Attribute & \cellcolor{Snow1!20}\xmark \\ \hline
    \cellcolor{SlateGray4!15}\textbf{Performance Efficiency} & \cellcolor{Snow1!20}\xmark & \cellcolor{Snow1!20}\xmark \\ \hline
    \cellcolor{SlateGray4!15}\textbf{Compatibility} & \cellcolor{Snow1!20}\xmark & \cellcolor{Snow1!20}\xmark \\ \hline
    \cellcolor{SlateGray4!15}\textbf{Usability} & \cellcolor{Snow1!20}Comment & \cellcolor{Snow1!20}\xmark \\ \hline
    \cellcolor{SlateGray4!15}\textbf{Reliability} & \cellcolor{Snow1!20}Ensure, File Mode, File & \cellcolor{Snow1!20}\xmark \\ \hline
    \cellcolor{SlateGray4!15}\textbf{Security} & \cellcolor{Snow1!20}SSH Key, File Mode & \cellcolor{Snow1!20}Hard-Coded String \\ \hline
    \cellcolor{SlateGray4!15}\textbf{Portability} & \cellcolor{Snow1!20}Comment & \cellcolor{Snow1!20}Lines of Code \\ \hline
    \cellcolor{SlateGray4!15}\textbf{Maintainability} & \cellcolor{Snow1!20}File & \cellcolor{Snow1!20}\xmark \\ \hline
    \end{tabular}
\bigskip

 \begin{tabular}{|l|}
    \hline
\rowcolor{SlateGray4!35}\multicolumn{1}{|c|}{\textbf{No Mapping}}    \\ \hline
    \cellcolor{Snow1!20}URL, Command \\ \hline
    \end{tabular}
    \vspace{0.5cm}\\
    \textbf{Legend:} \\
    \xmark \hspace{0.2cm} - No IaC code attributes maps to ISO 25010:2011 characteristics \\
\end{table}

\begin{table}[!ht]
    \centering
        \caption{Mapping of Code Quality Attributes of Dalla Palma et al.'s paper \cite{Palma2020TowardsAC} to ISO/IEC 25010:2011}
\label{mapping iso dalla palma}
    \begin{tabular}{|M{2cm}|M{2.5cm}|M{2.25cm}|}
    \hline
        \rowcolor{SlateGray4!35}\textbf{\textbf{\begin{tabular}[c]{@{}l@{}}ISO 25010:2011\\Characteristics\end{tabular}}} & \textbf{Direct Mapping} & \textbf{Indirect Mapping} \\ \hline
        \cellcolor{SlateGray4!15}\textbf{\begin{tabular}[c]{@{}l@{}}Functional\\Suitability\end{tabular}} & \cellcolor{Snow1!20}Include,Ensure & \cellcolor{Snow1!20}Variables \\ \hline
        \cellcolor{SlateGray4!15}\textbf{\begin{tabular}[c]{@{}l@{}}Performance\\Efficiency\end{tabular}} & \cellcolor{Snow1!20}Loops & \cellcolor{Snow1!20}\begin{tabular}[c]{@{}l@{}}Regex,\\Conditions,\\Decisions,\\Math Operation\end{tabular} \\ \hline
        \cellcolor{SlateGray4!15}\textbf{Compatibility} & \cellcolor{Snow1!20}\xmark & \cellcolor{Snow1!20}\xmark \\ \hline
        \cellcolor{SlateGray4!15}\textbf{Portability} & \cellcolor{Snow1!20}Paths,Import Playbook,Import Role,Include Role,Include Task,Import Task,Include Vars & \cellcolor{Snow1!20}Blocks \\ \hline
        \cellcolor{SlateGray4!15}\textbf{Reliability} & \cellcolor{Snow1!20}Ensure,File Mode,File,Error Handling,Ignore Errors & \cellcolor{Snow1!20}Blocks \\ \hline
        \cellcolor{SlateGray4!15}\textbf{Security} & \cellcolor{Snow1!20}SSH Key,File Mode & \cellcolor{Snow1!20}\begin{tabular}[c]{@{}l@{}}Suspicious\\Comments,\\Deprecated \\Keywords,\\Deprecated \\Modules\end{tabular} \\ \hline
        \cellcolor{SlateGray4!15}\textbf{Usability} & \cellcolor{Snow1!20}Comment & \cellcolor{Snow1!20}User Interactions \\ \hline
        \cellcolor{SlateGray4!15}\textbf{\begin{tabular}[c]{@{}l@{}}Maintainability \end{tabular}} & \cellcolor{Snow1!20}Comment & \cellcolor{Snow1!20}\begin{tabular}[c]{@{}l@{}}Line Blank,\\Line Source\\Code,\\Blank Space \\Between Words,\\Avg Play Size,\\Avg Task Size,\\Unique Names,\\Length of Tasks,\\Entropy,Blocks,\\Variable Names\end{tabular} \\\hline
    \end{tabular}
\bigskip

\begin{tabular}{|l|}
\hline
\rowcolor{SlateGray4!35}\multicolumn{1}{|c|}{\textbf{No Mapping}}                                                   \\ \hline
\cellcolor{Snow1!20}Play, Roles, URL, Commands, Fact Modules, Filters,\\\cellcolor{Snow1!20}Lookups, Keys, (External, Distinct) Modules \\ \hline
\end{tabular}
    \vspace{0.5cm}\\
    \textbf{Legend:} \\
    \xmark \hspace{0.2cm} - No IaC code attributes maps to ISO 25010:2011 characteristics 
\end{table}

\subsection*{STEP 2: DEVELOPMENT AND MAPPING OF IAC CODE ATTRIBUTES TO THE PROPOSED IaC CODE QUALITY FRAMEWORK}
\par
\scalebox{0.75}{\ding{108}} \textbf{RQ3:} \textit{Which specific IaC code quality categories can be developed to accommodate all identified IaC code attributes while drawing inspiration from ISO/IEC 25010:2011?}
\\

While ISO/IEC 25010:2011 offers a comprehensive model for software product quality, the previous analysis through the mappings shows that there is a need to adapt this standard to address the unique aspects of IaC. We propose a framework consisting of a set of quality categories that capture the uniqueness of IaC code as opposed to general software. This framework is compatible with the ISO standard as we show that the quality categories can be mapped to the the quality characteristics in the standard.

\begin{figure}[htbp!]
\centering
\includegraphics[width=\columnwidth]{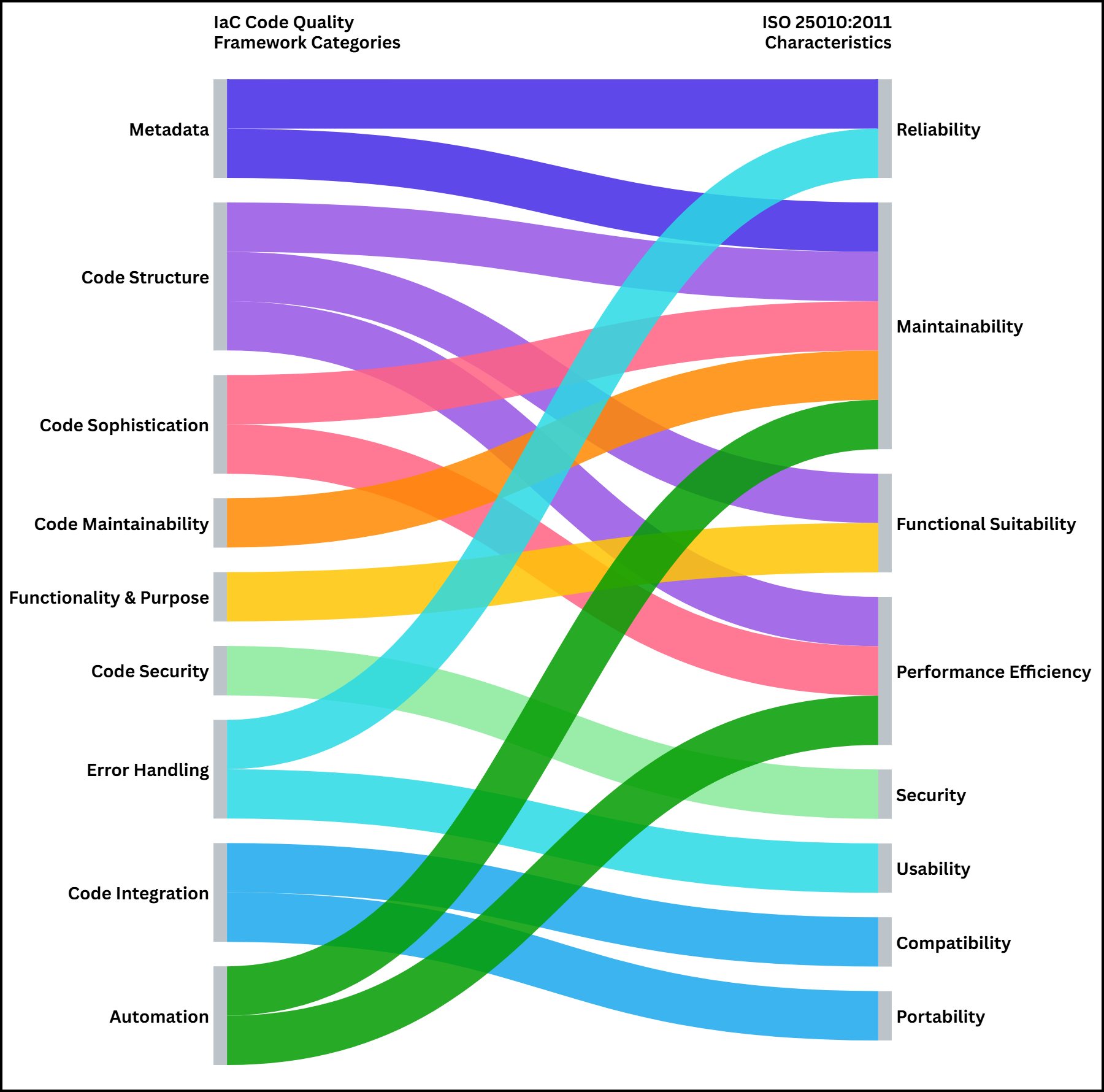}
\caption{Aligning of Our Proposed IaC Code Quality Framework to ISO/IEC 25010:2011 Characteristics}
\label{Figure:mapping_iso_to_iac}
\centering
\end{figure}

The IaC quality categories in our framework are \text{Metadata}, \text{Code Structure}, \text{Code Sophistication}, \text{Code Maintainability}, \text{Code Security}, \text{Function and Purpose}, \text{Error Handling}, \text{Automation}, and \text{Code Integration}. Fig.~\ref{Figure:mapping_iso_to_iac} shows these categories on the left hand side and also displays the relationship of each category with specific ISO/IEC 25010:2011 characteristics on the right hand side.
For instance, \text{Metadata} relates to maintainability and reliability by providing critical information about versioning, dependencies, and configuration parameters within the infrastructure code, which are vital for understanding and managing the code effectively.

The categories in the framework have been developed to emphasize the importance of well-organized and modular code (\text{Code Structure}), the use of advanced syntax (\text{Code Sophistication}), and the ease of modifying code to accommodate changes (\text{Code Maintainability}). It also highlights the importance of using security practices (\text{Code Security}), ensuring the code fulfills its intended functions (\text{Function and Purpose}), managing errors effectively (\text{Error Handling}), automating infrastructure processes to enhance efficiency (\text{Automation}), and enabling seamless integration across different environments (\text{Code Integration}). 

Using the same methodology as Rahman et al.'s \cite{Rahman_2019} and Dalla Palma et al.\ \cite{Palma2020TowardsAC}, we then mapped code quality attributes for each of the categories. To achieve this, data was collected from Ansible Galaxy repositories without factoring in popularity metrics like downloads or user ratings. This approach mitigated selection bias and allowed for an analysis that included a wide range of contributors. By incorporating metadata attributes not considered in earlier studies, such as those by Dalla Palma et al.\cite{Palma2020TowardsAC} and Begoug et al.\cite{TerraMetrics}, the analysis identified 95 unique Ansible-specific code attributes, contributing to a total of 104 attributes. These attributes were subsequently mapped to the proposed IaC code quality categories, as presented in Table~\ref{iac_code_metrics_to_framework}.

\begin{table}[!ht]
    \centering
    \caption{Summary of IaC Code Attributes Mapped to the Code Quality Framework Categories}
\label{iac_code_metrics_to_framework}
    \begin{tabular}{|p{1.70cm}|p{6.29cm}|}
    \hline
        \rowcolor{SlateGray4!35}\textbf{IaC Code Quality Category} & \centering\arraybackslash\textbf{IaC Code Attributes} \\ \hline
        \cellcolor{SlateGray4!15}\textbf{Metadata} & \cellcolor{Snow1!20}Download count, Tag, Total versions, Average update time, Dependencies, Supported platform count, Stars, Forks, Open issues, Minimum Ansible versions, version release time \\ \hline
        \cellcolor{SlateGray4!15}\textbf{\begin{tabular}[c]{@{}l@{}}Code\\Structure\end{tabular}} & \cellcolor{Snow1!20}Count of lines in YAML/YML files, Lines of code in YAML/YML files, Task count, Template files, Count of directories, Variables, Files count, Line blanks, Lines in source code, Blank spaces between words, Average play size, Average task size, Length of task, Unique names, Names with variables, Ensure \\ \hline
        \cellcolor{SlateGray4!15}\textbf{\begin{tabular}[c]{@{}l@{}}Code\\Sophistication\end{tabular}} & \cellcolor{Snow1!20}Entropy, Loops, Conditions, Decisions, Mathematical operations, Regex, Lookups, Include, Keys, External modules, Distinct modules, Fact modules \\ \hline
        \cellcolor{SlateGray4!15}\textbf{\begin{tabular}[c]{@{}l@{}}Code\\Maintainability\end{tabular}} & \cellcolor{Snow1!20}README count, README word count, License file presence, Version information, Entropy, Comments, User interactions \\ \hline
        \cellcolor{SlateGray4!15}\textbf{Functionality \& Purpose} & \cellcolor{Snow1!20}README count, Fact modules, Comments \\ \hline
        \cellcolor{SlateGray4!15}\textbf{\begin{tabular}[c]{@{}l@{}}Code\\Security\end{tabular}} & \cellcolor{Snow1!20}SELinux, Firewalld, apt\_key, passwd, Vault, stat, SSH keys, File modes, Suspicious comments, Deprecated keywords, Deprecated modules \\ \hline
        \cellcolor{SlateGray4!15}\textbf{\begin{tabular}[c]{@{}l@{}}Error\\Handling\end{tabular}} & \cellcolor{Snow1!20}Debug, failed\_when, changed\_when, Rescue, Always, Retry, Until loops, any\_errors\_fatal settings, max\_fail\_percentage, Delays, Error handling blocks, ignore\_errors, Blocks \\ \hline
        \cellcolor{SlateGray4!15}\textbf{Automation} & \cellcolor{Snow1!20}Become, vars, Handlers, Tags, check\_mode, environment, no\_log, local\_action, Fetch modules, Paths, Commands, URLs, Plays, Roles, External modules, Distinct modules, Fact modules, Filters \\ \hline
        \cellcolor{SlateGray4!15}\textbf{\begin{tabular}[c]{@{}l@{}}Code\\Integration\end{tabular}} & \cellcolor{Snow1!20}URI modules, URLs, wait\_for modules, Rsync usage, win\_service, add\_host, Git, Import playbooks, Import roles, Import tasks, Include roles, Include tasks, Include vars \\ \hline
    \end{tabular}
\end{table}

\subsection*{STEP 3: COMPARISON OF CURRENT RESEARCH MAPPING TO ISO/IEC 25010:2011 AND THE PROPOSED IaC CODE QUALITY FRAMEWORK}

In this subsection, we map the code attributes identified by Rahman et al.\ \cite{Rahman_2019}  and Dalla Palma et al.\ \cite{Palma2020TowardsAC}  onto our proposed IaC Code Quality Framework to evaluate its applicability. 
These mappings can be seen in  Table~\ref{mapping iac rahman} and Table \ref{mapping iac dalla}. Rahman et al.’s attributes were successfully incorporated into our \text{framework’s categories}, including \text{previously unmapped attributes} (see Table \ref{mapping_iso_rahman} \& Table \ref{mapping iso dalla palma}) of ``URL'' and ``Command,'' demonstrating the framework’s ability to integrate attributes from prior research. 
Similarly, all 46 quality attributes from Dalla Palma et al.\ \cite{Palma2020TowardsAC} for Ansible scripts can be mapped to our framework, illustrating its flexibility and applicability across different IaC tools and code quality assessments. This mapping process underscores the framework’s potential as a unifying model for IaC code quality evaluation.
Our IaC Code Quality Framework encompasses all attributes from both studies, aligning more closely with the unique aspects of IaC scripts and addressing elements not covered by general software quality models. This demonstrates that our framework is tailored to assessing IaC code quality. 

\begin{tcolorbox}[
    colframe=SlateGray4, 
    colback=Snow1!20,  
    boxrule=1pt, 
    rounded corners
]
\ding{228} \textbf{RQ3 Takeaway Message:}  \textit{Our IaC Code Quality Framework incorporates categories aligned with ISO/IEC 25010:2011 to encompass all identified IaC code attributes. It overcomes the limitations of the ISO standard in representing IaC-specific features, offering a holistic model for assessing IaC code quality.}
\end{tcolorbox}

\begin{table}[!ht]
    \centering
    \caption{Mapping Rahman et al.'s \cite{Rahman_2019} study to our proposed IaC Code Quality Framework}
\label{mapping iac rahman}
    \begin{tabular}{|M{2cm}|M{2cm}|M{1.25cm}|M{1.65cm}|}
    \hline
        \rowcolor{SlateGray4!35}\textbf{\begin{tabular}[c]{@{}l@{}}IaC Code\\Quality Category\end{tabular}} & \textbf{\begin{tabular}[c]{@{}l@{}}Direct\\Mapping \end{tabular}} & \textbf{\begin{tabular}[c]{@{}l@{}}Indirect\\Mapping \end{tabular}} & \textbf{\begin{tabular}[c]{@{}l@{}}No Mapping\\(Previously)\end{tabular}} \\ \hline
        \cellcolor{SlateGray4!15}\textbf{Metadata} & \cellcolor{Snow1!20}\xmark & \cellcolor{Snow1!20}\xmark & \cellcolor{Snow1!20}\xmark \\ \hline
        \cellcolor{SlateGray4!15}\textbf{Code Structure} & \cellcolor{Snow1!20}File & \cellcolor{Snow1!20}\begin{tabular}[c]{@{}l@{}}Attribute,\\Lines of\\ Code\end{tabular} & \cellcolor{Snow1!20}\xmark \\ \hline
        \cellcolor{SlateGray4!15}\textbf{\begin{tabular}[c]{@{}l@{}}Code\\Sophistication\end{tabular}} & \cellcolor{Snow1!20}\begin{tabular}[c]{@{}l@{}}Include,Ensure,\\Require\end{tabular} & \cellcolor{Snow1!20}\xmark & \cellcolor{Snow1!20}\xmark \\ \hline
        \cellcolor{SlateGray4!15}\textbf{\begin{tabular}[c]{@{}l@{}}Code\\Maintainability\end{tabular}} & \cellcolor{Snow1!20}Comment & \cellcolor{Snow1!20}\xmark & \cellcolor{Snow1!20}\xmark \\ \hline
        \cellcolor{SlateGray4!15}\textbf{\textbf{\begin{tabular}[c]{@{}l@{}}Functionality\\\& Purpose\end{tabular}}} & \cellcolor{Snow1!20}Comment & \cellcolor{Snow1!20}\xmark & \cellcolor{Snow1!20}\xmark \\ \hline
        \cellcolor{SlateGray4!15}\textbf{Code Security} & \cellcolor{Snow1!20}SSH Key,File Mode & \cellcolor{Snow1!20}Hard-Coded String & \cellcolor{Snow1!20}\xmark \\ \hline
        \cellcolor{SlateGray4!15}\textbf{Error Handling} & \cellcolor{Snow1!20}\xmark & \cellcolor{Snow1!20}\xmark & \cellcolor{Snow1!20}\xmark \\ \hline
        \cellcolor{SlateGray4!15}\textbf{Automation} & \cellcolor{Snow1!20}\xmark & \cellcolor{Snow1!20}\xmark & \cellcolor{Snow1!20}\begin{tabular}[c]{@{}l@{}}URL,\\Commands\end{tabular} \\ \hline
        \cellcolor{SlateGray4!15}\textbf{Code Integration} & \cellcolor{Snow1!20}\begin{tabular}[c]{@{}l@{}}Include,Ensure,\\Require\end{tabular} & \cellcolor{Snow1!20}\xmark & \cellcolor{Snow1!20}\xmark \\ \hline
    \end{tabular}
    \vspace{0.5cm}\\
    \textbf{Legend:} \\
    \xmark \hspace{0.2cm} - No IaC code attributes maps to our proposed IaC code quality category \\
\end{table}

\begin{table}[!ht]
    \centering
\caption{Mapping Dalla Palma et al.'s \cite{Palma2020TowardsAC} study to our proposed IaC Code Quality Framework}
\label{mapping iac dalla}
    \begin{tabular}{|M{1.95cm}|M{1.75cm}|M{1.80cm}|M{1.50cm}|}
    \hline
        \rowcolor{SlateGray4!35}\textbf{\begin{tabular}[c]{@{}l@{}}IaC Code\\Quality Category\end{tabular}} & \textbf{\begin{tabular}[c]{@{}l@{}}Direct\\Mapping \end{tabular}} & \textbf{\begin{tabular}[c]{@{}l@{}}Indirect\\Mapping \end{tabular}} & \textbf{\begin{tabular}[c]{@{}l@{}}No Mapping \\ (Previously) \end{tabular}} \\ \hline
        \cellcolor{SlateGray4!15}\textbf{Metadata} & \cellcolor{Snow1!20}\xmark & \cellcolor{Snow1!20}\xmark & \cellcolor{Snow1!20}\xmark \\ \hline
        \cellcolor{SlateGray4!15}\textbf{Code Structure} & \cellcolor{Snow1!20}File & \cellcolor{Snow1!20}Variables,Lines Blank,Lines Source Code,Blank Space Between Words,Avg Play Size,Avg Task Size,Length Tasks,Unique Names,Names with Variables & \cellcolor{Snow1!20}\xmark \\ \hline
        \cellcolor{SlateGray4!15}\textbf{\begin{tabular}[c]{@{}l@{}}Code\\Sophistication\end{tabular}} & \cellcolor{Snow1!20}Include,Loops & \cellcolor{Snow1!20}\begin{tabular}[c]{@{}l@{}}Entropy,\\Conditions,\\Decisions,\\Math Operations,\\Regex\end{tabular} & \cellcolor{Snow1!20}\begin{tabular}[c]{@{}l@{}}Lookups,Keys,\\(External,\\ Fact, Distinct)\\Modules\end{tabular}
 \\ \hline
        \cellcolor{SlateGray4!15}\textbf{\begin{tabular}[c]{@{}l@{}}Code\\Maintainability\end{tabular}} & \cellcolor{Snow1!20}Comment & \cellcolor{Snow1!20}Entropy,User Interaction & \cellcolor{Snow1!20}\xmark \\ \hline
        \cellcolor{SlateGray4!15}\textbf{\begin{tabular}[c]{@{}l@{}}Functionality\\\& Purpose\end{tabular}} & \cellcolor{Snow1!20}Comment & \cellcolor{Snow1!20}\xmark & \cellcolor{Snow1!20}\xmark \\ \hline
        \cellcolor{SlateGray4!15}\textbf{Code Security} & \cellcolor{Snow1!20}SSH Key,File Mode & \cellcolor{Snow1!20}\begin{tabular}[c]{@{}l@{}}Suspicious \\Comments,\\Deprecated\\Keywords,\\Deprecated\\Modules\end{tabular} & \cellcolor{Snow1!20}\xmark \\ \hline
        \cellcolor{SlateGray4!15}\textbf{Error Handling} & \cellcolor{Snow1!20}Error Handling Blocks,Ignore Errors & \cellcolor{Snow1!20}Blocks & \cellcolor{Snow1!20}\xmark \\ \hline
        \cellcolor{SlateGray4!15}\textbf{Automation} & \cellcolor{Snow1!20}Paths & \cellcolor{Snow1!20}\xmark & \cellcolor{Snow1!20}\begin{tabular}[c]{@{}l@{}}Commands,\\URL, Filters,\\Play,Roles,\\(External,\\Distinct, Fact)\\ Modules\end{tabular} \\ \hline
        \cellcolor{SlateGray4!15}\textbf{Code Integration} & \cellcolor{Snow1!20}Import Playbook,Import Role,Import Task,Include Role,Include Task,Include Vars & \cellcolor{Snow1!20}\xmark & \cellcolor{Snow1!20}\xmark \\ \hline
    \end{tabular}
    \vspace{0.5cm}\\
    \textbf{Legend:} \\
    \xmark \hspace{0.2cm} - No IaC code attributes maps to our proposed IaC code quality category \\
\end{table}

\section{COMPUTATION OF IaC CODE ATTRIBUTES}

In this section, we explain how we computed the code attributes for IaC. We began by counting the frequency of each code attribute and then adjusted these metrics per 100 lines of code, denoted by \( \text{m}_{100} \), allowing comparisons between repositories of different sizes. Subsequently, we normalized the metrics to the range 0--1, denoted by (\( \text{nm}_{100} \)), which facilitates direct comparison by placing all metrics on a common scale, regardless of their original units or magnitudes.

Some tabulated metrics represent negative measures. For instance, higher counts for Suspicious Comments, Use of \textit{passwd}, Deprecated keywords, and Deprecated modules indicate poorer code quality.  
For such metrics, their normalisation values were inverted using Equation~\ref{eq:nm100_neg}.
In doing so, this meant all the calculated metrics were in a form independent of repository size and aligned to the scales for consistent interpretation and aggregation.

\begin{equation}\label{eq:nm100_neg}
\tcbhighmath{\text{nm}_{100,\text{neg}} = 1 - \frac{m_{100,\text{neg}}}{\max(m_{100,\text{neg}})}}
\end{equation}

Overall category scores per 100 lines of code were calculated by aggregating the normalized positive and negative metrics, weighted according to their importance, using the equation:
\begin{equation}\label{eq:cs100}
\tcbhighmath[colback=Ivory1!0, colframe=Snow1!0, boxrule=0.5pt,width=auto, right=0.5pt, bottom=0.01pt, coltext=black]{\text{cs}_{100} = \frac{\sum_{i=1}^{m} (\text{nm}_{100,i} \times w_{i}) + \sum_{j=1}^{n} (\text{nm}_{100,\text{neg}, j} \times w_{j})}{N}},
\end{equation}
where \( \text{nm}_{100} \) and \( \text{nm}_{100,\text{neg}} \) are the normalized results of positive and negative metrics, respectively, \( w_{i} \) and \( w_{j} \) are the weights assigned to each positive and negative attribute,
and \( N = m + n \) is the total number of positive (\( m \)) and negative (\( n \)) attributes incorporated in the calculation. The total score was derived by summing the category scores across all nine categories, resulting in a score ranging from 0--9, as each category score ranges from 0--1. This methodology ensured that each repository's code quality was assessed uniformly, accounting for size and the significance of each metric.

To measure the \textit{Total Score}, the category scores per 100 lines of code are summed across all nine categories into a single, interpretable score using the following equation:

\begin{equation}\label{eq:total_score}
\tcbhighmath{\text{Total Score} = \sum_{i=1}^{9} (\text{cs}_{100,i})},
\end{equation}
where \( \text{cs}_{100,i} \) represents the category score per 100 lines of code for each category in the proposed IaC framework, and \textit{Total Score} denotes the aggregate score per 100 lines of code.


The assignment of weights determines the importance of attributes in code quality assessment. Weights (\( w_{i} \) \& \( w_{j} \) as shown in Equation \ref{eq:cs100}) ranging from 0--1 allow for the prioritization of attributes, influencing the overall quality score. This flexibility allows organizations to:

\begin{itemize}
    \item Measure code quality in accordance with software development practices and policies.
    \\
    \item Encode policies into quality measurement by assigning weights to attributes. 
\end{itemize}

In our study, a weighting approach assigning equal importance to all attributes was employed to avoid bias in the evaluation. Utilizing a weighted average in the scoring model enables:

\begin{itemize} 
    \item Adjustments in response to changes in organizational policies.
    \\
    \item Maintenance of evaluation relevance as priorities evolve. 
\end{itemize}



\section{TEMPORAL DYNAMICS OF CODE QUALITY}
\par
\scalebox{0.75}{\ding{108}} \textbf{RQ4:} \textit{How does the proposed IaC code quality framework perform when applied to real-world IaC repositories, and what insights can be gained?}
\\

\begin{figure}[htbp!]
\centering
\includegraphics[width=1.05\columnwidth]{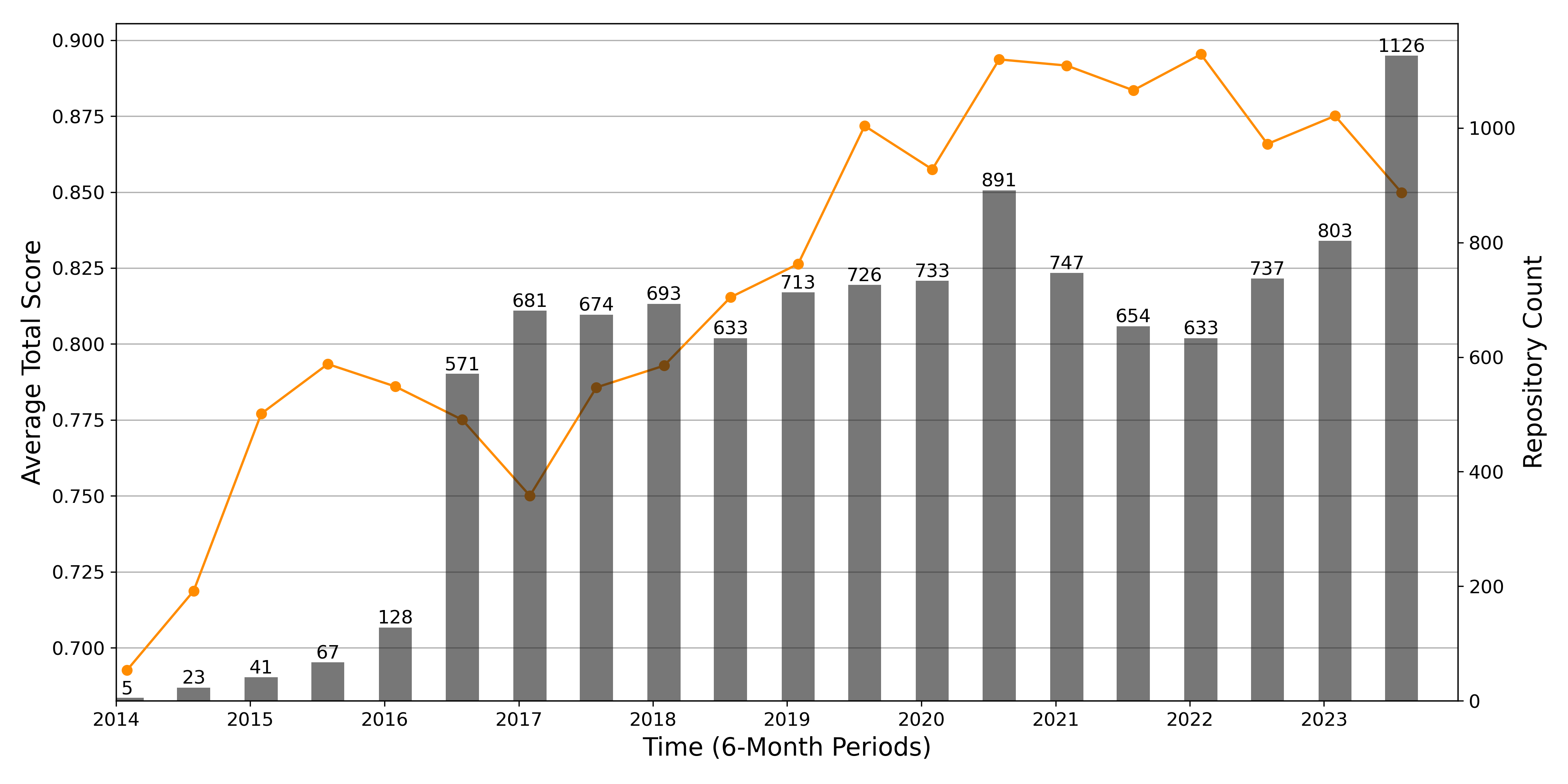}
\caption{6-Month Averages of Total Score Across Time}
\label{Figure:fulldataset_timeline}
\centering
\end{figure}

\begin{figure*}[htbp!]
\centering
\includegraphics[width=\textwidth]{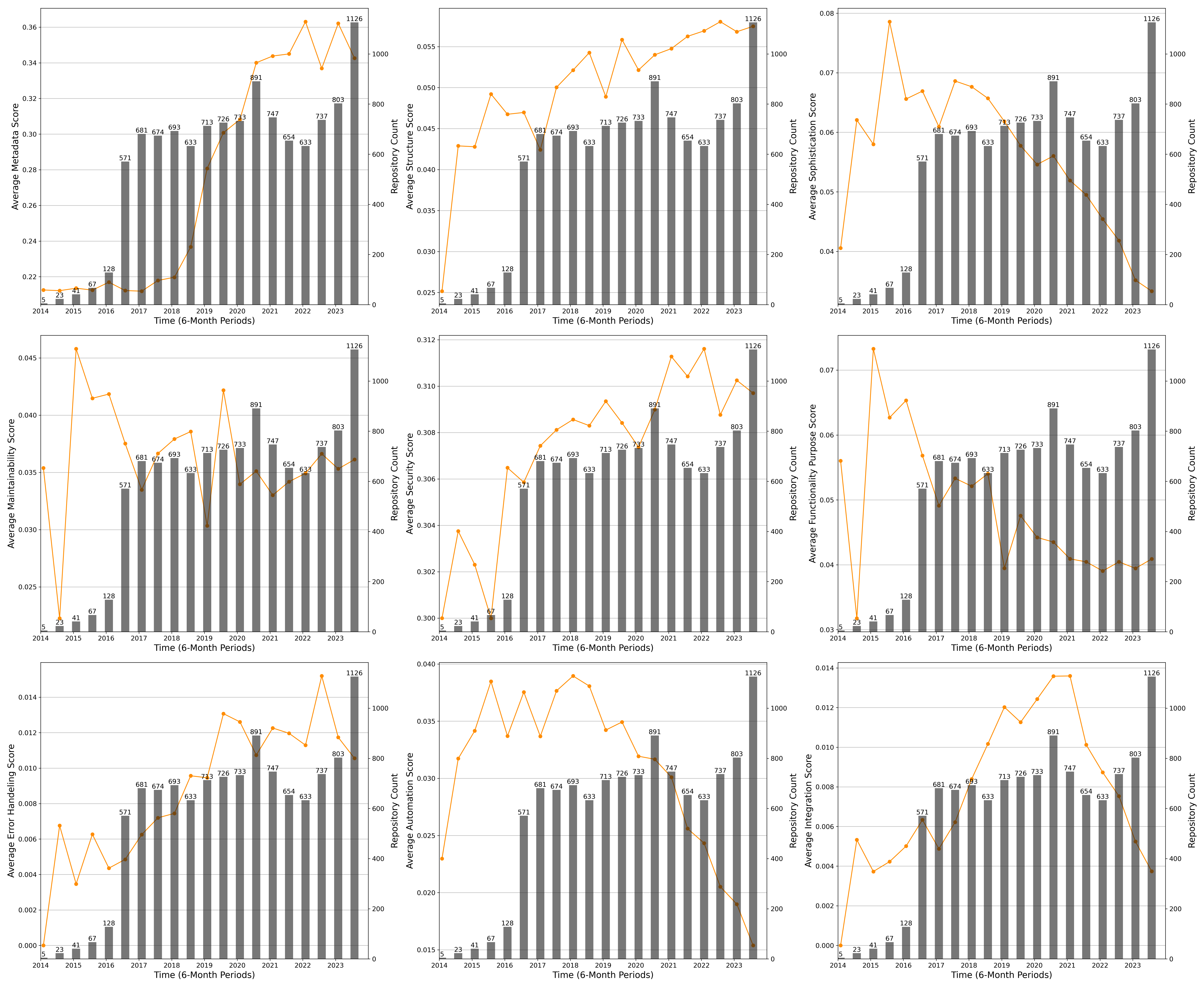}
\caption{6-Month Average of Category Scores Across Time}
\label{Figure:fulldataset_trends_indetail}
\centering
\end{figure*}

In this section, we apply our framework on IaC scripts and present two primary observations: the \text{time series analysis} and an \text{assessment of observable trends} overtime of the quality of IaC and specifically Ansible scripts. This analysis uses data from \text{11,279} Ansible Galaxy repositories with complete records to evaluate code quality across various IaC practices within the Ansible community. The time series graph illustrates changes in \text{mean scores} over time, along with repository counts for each \text{six-month period}. Using a dual-axis layout, the graphs in Fig.~\ref{Figure:fulldataset_timeline} and Fig.~\ref{Figure:fulldataset_trends_indetail} facilitate a direct comparison between repository quality (mean scores) and repository counts over time. Each bar is labeled with the repository count, allowing for an examination of changes in dataset composition. The mean scores for each interval reflects the number of repositories included, which varies according to the dataset composition for that period. Fig.~\ref{Figure:fulldataset_timeline} illustrates this for total score over all categories while Fig.~\ref{Figure:fulldataset_trends_indetail} shows individual categories.

This longitudinal analysis identifies trends in repository quality and quantity, which may suggest changes in practices within the Ansible community. Using this methodology, we can examine how evolution influences mean scores over time, providing insights into broader IaC patterns within the Ansible Galaxy ecosystem. An initial review of the graphs reveals an upward trend in scores; however, further analysis indicates that the scores remain independent of repository count over time. One notable trend observed was an increase in average total score as shown in Fig.~\ref{Figure:fulldataset_timeline} during the period from mid-2020 to mid-2022, corresponding with the COVID-19 pandemic \cite{10.1145/3419394.3423658}. This period saw a global shift towards online services, leading many developers to prioritize IaC, with a focus on enhancing IT system resilience and security. This increase in repository quality during the pandemic years contrasts with data from the broader 2014--2023 period, highlighting a trend of increased code quality and attention to IaC standards in response to heightened demand for robust digital infrastructure. 

\begin{figure}[htbp!]
\centering
\includegraphics[width=\columnwidth]{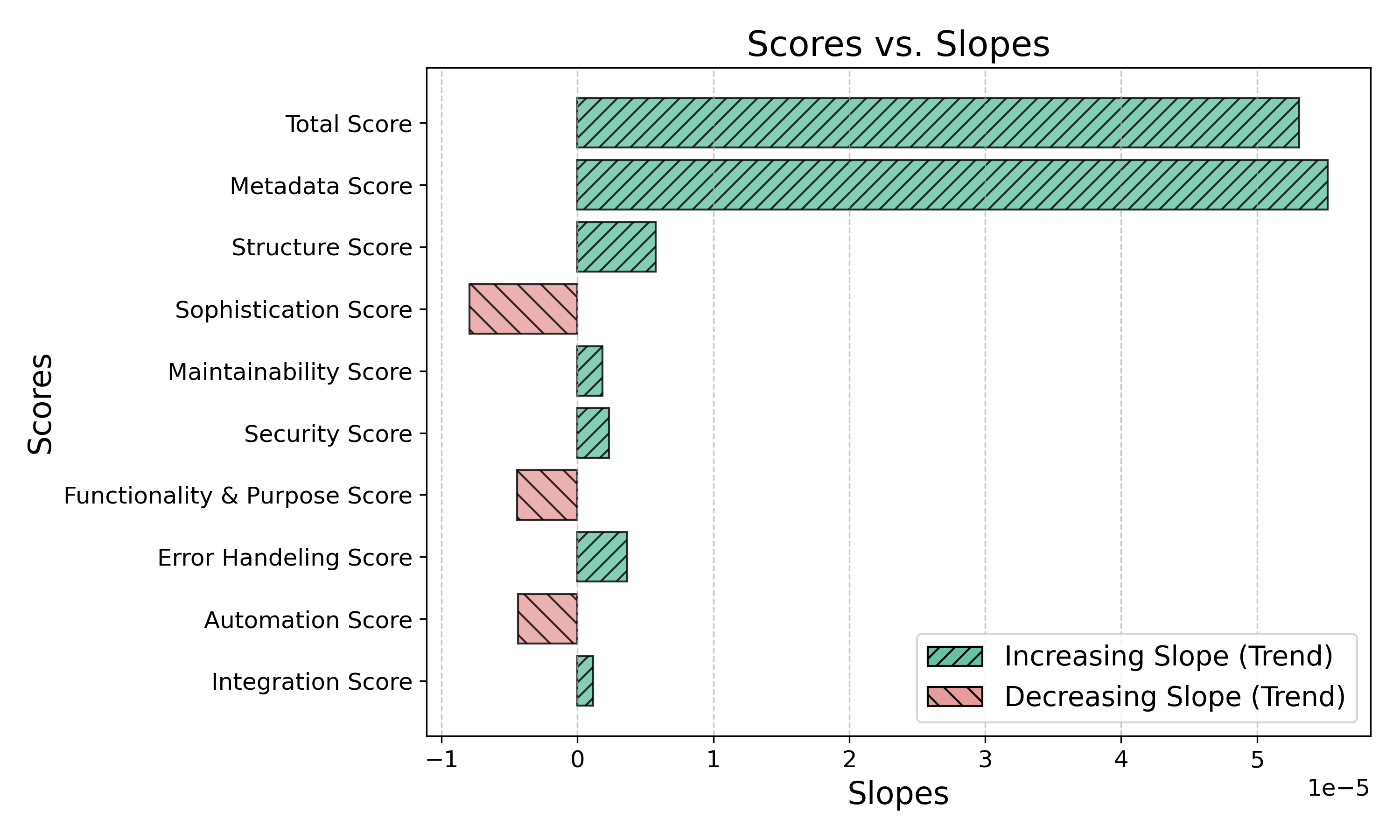}
\caption{Observable Trends}
\label{Figure:fulldataset_trends}
\centering
\end{figure}

Lastly, an Ordinary Least Squares (OLS) regression \cite{stock2020introduction} was performed with six-month intervals as the independent variable and the mean scores as the dependent variable from the data shown in the graphs Fig.~\ref{Figure:fulldataset_timeline} and Fig.~\ref{Figure:fulldataset_trends_indetail}. Regression parameters, where slope values reveal trends over time as illustrated in Fig.~\ref{Figure:fulldataset_trends}.

Notably, the \textit{Total Score} exhibited a consistent increasing trend which indicates a steady improvement in overall code quality over time. Similarly, the \textit{Metadata Score} demonstrated the most significant increasing trend which reflects user's growing emphasis on the comprehensive metadata to enhance repository usability. Increasing trends were also identified in the \textit{Structure Score} and \textit{Security Score}, suggesting improvements in organizational structure and increased attention to security practices to align with best practices in the IaC industry. Additionally, the \textit{Error Handling Score} showed a significant increase which indicates a growing focus on error management and self-healing capabilities.

Conversely, certain categories exhibited declining trends. The \textit{Sophistication Score} showed a decreasing trend, suggesting a decrease in the use of diverse keywords in repositories. The \textit{Functionality \& Purpose Score} also demonstrated a statistically significant decreasing trend, indicating a potential shift towards simpler and more focused repositories. Similarly, the \textit{Automation Score} decreased over time, indicating a decline in the usage of automation-specific keywords. This is likely due to the improvements in the base Ansible-core, which gradually incorporates more advanced keywords from various third-party modules.

Overall, these trends highlight a community-wide focus on improving code quality in areas such as metadata completeness, structural organization, security, and error handling. The decline in usage of attributes related to sophistication, automation, and functionality may reflect a shift towards simplicity and maintainability, indicating evolving priorities within the community towards more streamlined IaC practices.

\begin{tcolorbox}[
    colframe=SlateGray4, 
    colback=Snow1!20,  
    boxrule=1pt, 
    rounded corners
]
\ding{228} \textbf{RQ4 Takeaway Message:} \textit{Using our IaC code quality framework to analyze real-world Ansible repositories demonstrated overall improvements in code quality and offered insights into evolving practices within the ansible community.}
\end{tcolorbox}

\section{DISCUSSION}
\par
\scalebox{0.75}{\ding{108}} \textbf{RQ5:} \textit{In what ways can the proposed IaC code quality framework enhance the assessment of IaC scripts in practice?}
\\

The proposed IaC code quality framework evaluates Ansible scripts by establishing definitions backed by standardization, ensuring consistent assessments across repositories and authors. \text{Practitioners} can benchmark their code against industry standards using quantifiable metrics that identify strengths and weaknesses. The framework emphasizes the importance of detailed metadata for enhancing repository discoverability and usability, providing insights for tracking progress and adjusting strategies. \text{For researchers}, the framework provides a basis for comparative studies across IaC tools, the incorporation of additional attributes and metrics, and the utilization of large-scale repository data for quantitative analysis of IaC development practices. Researchers can examine relations between specific metrics and overall quality or explore deeper connections using various data analysis techniques.

\subsection{LIMITATIONS AND RECOMMENDATIONS}

This study identifies key limitations and proposes recommendations for future research.

\scalebox{0.75}{\ding{110}} \textbf{Limitations:}
\begin{itemize}
    \item \textbf{Metadata Completeness Criterion}
    \begin{itemize}
        \item The analysis included 11,279 repositories out of approximately 26,900 in Ansible Galaxy (42\%) due to the requirement for complete metadata.
        \item This selection criterion exclude repositories with incomplete metadata, potentially introducing bias and omitting repositories with distinct characteristics.
    \end{itemize}

    \item \textbf{Tool-Specific Scope}
    \begin{itemize}
        \item The study focused exclusively on Ansible repositories from Ansible Galaxy, which limits the applicability of findings to other IaC tools.
    \end{itemize}
    
    \item \textbf{Framework Weighting Influence}
    \begin{itemize}
        \item The assigned weights in the framework metrics can affect the interpretation of their importance, necessitating thorough testing and careful evaluation to mitigate bias.

    \end{itemize}

    \item \textbf{Correlations Analysis}
    \begin{itemize}
        \item While trends were identified, our study needs to establish clear correlations between specific code attributes and category-specific quality scores.

    \end{itemize}
\end{itemize}

\scalebox{0.75}{\ding{110}} \textbf{Recommendations:}
\begin{itemize}
    \item Expand the scope to include other IaC tools and repository hosting platforms.
    \item Refine the relationship between organizational policies and assigned weights based on empirical evidence or expert feedback.
    \item Incorporate qualitative research methods, such as:
    \begin{itemize}
        \item Conducting surveys or interviews with developers to complement quantitative analyses.
        \item Gaining insights into coding practices and community norms.
    \end{itemize}
    \item Adjust metric definitions and experiment with weightings to align with current best practices.
    \item Identifying and categorizing outliers, whether beneficial or detrimental, remains a critical aspect of data analysis that warrants further investigation.
\end{itemize}

\begin{tcolorbox}[
    colframe=SlateGray4, 
    colback=Snow1!20,  
    boxrule=1pt, 
    rounded corners
]
\ding{228} \textbf{RQ5 Takeaway Message:} \textit{Our study provides metrics for practitioners to assess script quality and offers researchers a basis for deeper exploration of IaC development practices.}
\end{tcolorbox}

\section{CONCLUSION}

This paper has focused on the growing trend of IaC for automated configuration through scripting, highlighting the lack of standardized code quality evaluation aligned with a software quality standard such as ISO/IEC 25010:2011. To bridge this gap, we have proposed an IaC quality framework for evaluating Ansible scripts using definitions aligned with standards for consistent assessments across repositories. In this process, we identified 95 unique Ansible-specific code attributes. With it practitioners, for instance, can benchmark their code against industry standards using quantifiable metrics to identify strengths and weaknesses. Emphasizing metadata importance for repository usability, we have formulated nine IaC specific code quality categories. 
Taking a decade's worth of Anisble-based repositories, analysis reveals a trend towards
enhancing code quality in areas like metadata completeness, structural organization, security, and error handling. A decreased emphasis on sophistication, automation, and functionality was also noted, which suggests a shift towards simplicity of code, indicating evolving priorities toward more streamlined IaC practices.

\section{DATA AVAILABILITY}

The data supporting the findings of this study are openly and anonymously available on Figshare at: \textit{https://figshare.com/s/b78d2c82767ab7d86beb}

\section*{Acknowledgment}

We appreciate the valuable feedback provided by the anonymous reviewers. The authors would also like to acknowledge funding support from the New Zealand Ministry of Business, Innovation, and Employment (MBIE) for project UOWX1911, ``Artificial Intelligence for Human-Centric Security”.

\clearpage
\IEEEtriggeratref{16}

\bibliographystyle{IEEEtran}
\bibliography{IEEEabrv,ref}

\end{document}